\documentclass[conference]{IEEEtran}
\usepackage{cite}

\ifCLASSINFOpdf
 \usepackage[pdftex]{graphicx}
\else
 \usepackage[dvips]{graphicx}
\fi
 \usepackage{xspace}
 \newcommand{\cp}{{\sf \sc Commet}\xspace}
\usepackage{sidecap}
\usepackage{amssymb}
\newcommand{\oc}{\mbox{ $\cap$\protect\raisebox{.6em}{\kern-0.8em$\leadsto$}}\xspace}
\newcommand{\ocf}{\mbox{ $\Cap$\protect\raisebox{.6em}{\kern-0.8em$\leadsto$}}\xspace}
\begin{document}
\bstctlcite{IEEEexample:BSTcontrol}
%
\title{\cp: comparing and combining multiple metagenomic datasets}



%
\author{\IEEEauthorblockN{Nicolas Maillet\IEEEauthorrefmark{1}, Guillaume Collet\IEEEauthorrefmark{2}, Thomas Vannier\IEEEauthorrefmark{3}, Dominique Lavenier\IEEEauthorrefmark{1} and Pierre Peterlongo\IEEEauthorrefmark{1}$^\divideontimes$}
\IEEEauthorblockA{\IEEEauthorrefmark{1}INRIA / IRISA-UMR CNRS 6074, EPI GenScale, Rennes, France\\}
\IEEEauthorblockA{\IEEEauthorrefmark{2}INRIA / IRISA-UMR CNRS 6074, EPI Dyliss, Rennes, France\\}
\IEEEauthorblockA{\IEEEauthorrefmark{3}CEA Genoscope / CNRS UMR 8030 / Universit\'e d'\'Evry, Evry, France}%
{$^\divideontimes$ Corresponding author: Pierre Peterlongo pierre.peterlongo@inria.fr}
}


\maketitle

\begin{abstract}

Metagenomics offers a way to analyze biotopes at the genomic level and to reach functional and taxonomical conclusions. 
The bio-analyzes of large metagenomic projects face critical limitations: complex metagenomes cannot be assembled and the taxonomical or functional annotations are much smaller than the real biological diversity. 
This motivated the development of \emph{de novo} metagenomic read comparison approaches to extract information contained in metagenomic datasets.

However, these new approaches do not scale up large metagenomic projects, or generate an important number of large intermediate and result files.
We introduce \cp (``COmpare Multiple METagenomes''), a method that provides similarity overview between all datasets of large metagenomic projects.

Directly from non-assembled reads, all against all comparisons are performed through an efficient indexing strategy.
Then, results are stored as bit vectors, a compressed representation of read files, that can be used to further combine read subsets by common logical operations.
Finally, \cp computes a clusterization of metagenomic datasets, which is visualized by dendrogram and heatmaps.

Availability: http://github.com/pierrepeterlongo/commet
\end{abstract}


%
\IEEEpeerreviewmaketitle

\section{Introduction}
\label{introduction}
NGS revolution enabled the emergence of the metagenomic field where an environment is sequenced instead of an individual or a species, opening the way to a comprehensive understanding of environmental microbial communities. 
Large metagenomic projects such as MetaSoil~\cite{Delmont2012}, MetaHit~\cite{Qin2010} or Tara Oceans~\cite{Karsenti2011} witness this evolution. 
Analyzing of metagenomic data is a major bottleneck. 
For instance, assembly tests over ``simple'' simulated metagenomes showed that N50 is only slightly larger than read sizes~\cite{Pignatelli2011}. 
This situation becomes even worth on complex datasets, such as seawater, where millions of distinct species coexist.
In this case, biodiversity can be estimated by using statistical approaches~\cite{Wang2012} or by mapping reads on reference banks~\cite{Huson2007,Markowitz2012}.
Nevertheless, statistical approaches are limited to a few dozens of species with limited differences in their relative abundance. In addition, the mapping approaches are limited to current knowledge contained in reference banks that suffer from their incompleteness and their inherent errors~\cite{Schnoes2009}.

A key point of substantial metagenomic projects stands in the number of metagenomes they produce.
Then, similarities and differences between metagenomes can be exploited as a source of information, measuring external effects like pollution sources, geographic locations, and patient microbial gut environment~\cite{Qin2010,Delmont2012b}.
A few methods were proposed to compare metagenomes using external information sources such as taxonomic diversity~\cite{Jaenicke2011} or functional content~\cite{Sommer2009}. However, these methods are biased because as they are based on partial knowledge.

Methods were proposed to compare metagenomes without using any \emph{a priori} knowledge.
These \emph{de novo} methods use global features like $GC$ content~\cite{Foerstner2005}, genome size~\cite{Raes2007} or sequence signatures~\cite{Jiang2012}. 
Theses methods face limitations as they are based on rough imprecise criteria and as they only compute a similarity distance: they do not extract similar elements between samples.
We believe that it is possible to go further by comparing metagenomic samples at the read sequence level. This provides a higher precision distance and, importantly, it provides reads that are similar between datasets or that are specific to a unique dataset, enabling their latter analysis: assembly with better coverage or comparison with other metagenomic samples.
Such comparisons may be performed using Blast~\cite{Altschul1990403} or Blat~\cite{Kent2002} like tools.
Unfortunately, these methods do not scale up on large comparative metagenomic studies in which hundreds of millions of reads have to be compared to other hundreds of millions of reads. 
For instance, one can estimate that comparing a hundred of metagenomes each composed by a hundred of millions of reads of size 100 would require centuries of CPU computation. 
The crAss approach~\cite{Dutilh2012} constructs a reference metagenome by cross assembling reads of all samples. 
Then, it maps the initial reads on the so obtained contigs and several measures are derived, based on the repartition of mapped reads. 
This method provides results of high quality. 
However, due to its assembly and mapping approach, it does not scale up to large metagenomic datasets.
Simpler methods such as TriageTools~\cite{Fimereli2013} or Compareads~\cite{Maillet2012} measure the sequence similarity of a read with a databank by counting the number of $k$-mers (words of length $k$) shared with the databank. 
Due to memory consumption, TriageTools cannot use $k$ values larger than 15 and is thus limited to small datasets (a few hundred of thousands reads of length 100).
The Compareads tool scales up to large datasets with a small memory footprint and acceptable running time.
However, applied on large metagenomic projects, this tools generates an important number of large intermediate result files. 
In practice, applying Compareads to $N$ datasets generates $N^2$ resulting new datasets, each of the size of the original ones at worse. 
Additionally, Compareads leads to highly redundant computation raising up the execution time. 
These drawbacks are serious bottlenecks limiting the practical usage of Compareads. 

In this paper, we introduce \cp (``COmpare Multiple METagenomes''), a fast software that provides a global similarity overview between all datasets of a metagenomic project.
\cp is based on the Compareads philosophy that consists in determining similarity between two metagenomic datasets by extracting common reads using $k$-mer approach: two reads are considered similar if they share $t$ non-overlapping $k$-mers ($t$ and $k$ are parameters). 
A metagenomic project involving $N$ datasets will thus requires the computation of $N^2$ intersections which is both time- and storage-consuming. 
To keep computation time as low as possible, the computation of the $N^2$ intersections has been strongly improved compared to the Compareads approach through an efficient indexing strategy in which each file is fully indexed only once. 
In addition, to save storage space, intersections between metagenomic datasets are represented as bit vectors.
This compact representation reduces the storage space by two orders of magnitude. 
Moreover, it provides an easy way to filter and sub-sample reads, or to combine various results by applying logical operations. 
Finally, \cp computes a clusterization of metagenomic datasets, which is visualized by dendrogram and heatmaps.

\section{Method}

\subsection{Comparing two sets of reads} 
\label{sub:two_sets}

The \cp algorithm to compare two sets of read is based on the Compareads~\cite{Maillet2012} methodology. 
It consists in finding reads from a set $A$ that are similar to at least one read from a set $B$. 
The similarity between two reads is based on a minimal number $t$ of non-overlapping identical $k$-mers.
This core operation is directed : it provides reads from $A$ similar to reads from $B$ but it does not provides reads from $B$ similar to reads from $A$. 
Note that, as explained below, this operation is based on a heuristic.  
Thus we denote this operation by $A\oc B$.

Computing $A\oc B$ consists in two steps.
Firstly, $k$-mers from $B$ are indexed in a Bloom filter like data-structure~\cite{Bloom1970}.
Secondly, non-overlapping $k$-mers of reads from $A$ are searched in the Bloom filter.
A read $r$ from $A$ sharing $t$ non-overlapping $k$-mers with the Bloom filter is considered similar to at least one read from $B$.
However, the algorithm does not check that these $k$-mers co-occur on a single read from $B$, which is a source of false positives. 
Readers are invited to refer to~\cite{Maillet2012} for having more details on precision results.

We recall that the following strategy is applied in order to limit the second source of false positives.
First $A \oc B$ is computed. 
Then, instead of naïvely computing $B \oc A$, $B\oc (A\oc B)$ is computed. 
This limits the indexed reads of $A$ to those already detected as similar to at least one read from $B$. 
Finally, the symmetrical operation is performed: $A\oc(B\oc(A\oc B))$.

The previously exposed strategy to fully compare sets $A$ and $B$ within three consecutive $\oc$ operations has also the advantage to limit the indexation effort. 
Indeed, only the first $A\oc B$ operation indexes the full set $B$. 
The two other operations only index subsets of $A$ and $B$.

While comparing read samples $A$ and $B$, the final results of interest are the reads of $A$ similar to reads of $B$ computed by $A\oc(B\oc(A\oc B))$ and reads of $B$ similar to reads of $A$ computed by $B\oc (A\oc B)$. For sake of simplicity, we denotes these two sets as, respectively, $A\ocf B$ and $B\ocf A$.

In the following sections we present the \cp novelties: represent read subsets with a limited disk space impact, new read filtering and read subsets manipulation features, compare multiple sets of reads, visualize dataset's similarities as heatmaps and dendrogram.

\subsection{Read subsets representation} 
\label{sub:sub_read_sets_representation}
In \cp we propose a simple yet compact data structure to represent a read subset: a vector of bits where each bit represents a read of the original read set. 
This is what we call the "bit vector representation".
As shown below, this representation enables to filter and to subsample read files, to represent $\oc$ (and thus $\ocf$) results and to easily perform logical operation between read subsets.

Note that with such a representation, a bit vector needs hundreds to thousand times less disk space than a classical uncompressed fastq file.
Note also that this way of coding read subsets is not limited to the \cp framework. 
It may be applied to any other programs that manipulate read subsets. 
Thus, the \cp tool includes a C++ library of reusable components to manipulate read subsets.

In the \cp framework, the bit vector representation is used as inputs and/or outputs of all tools. In particular they are used in the following operations:
\subsubsection{Read subsampling and filtering} 
\label{ssub:read_subsampling_and_read_filtering}

With huge datasets, it may appear necessary to subsample, for instance limiting each read file to a same number $m$ of a few millions reads. 
This is immediate by creating a bit vector in which only the first $m$ bits are set to $1$, while others are set to $0$.

Raw NGS reads also usually need to be filtered on several practical characteristics (read size, read complexity, \dots).
Thus, a bit vector is a direct representation of a filtered result: bit values associated to selected reads are set to $1$, the others to $0$.
A combination of subsampling and filtering allows to select only the $m$ first reads that fulfill the filtration criteria.

\subsubsection{Representing the similar reads} 
\label{ssub:representing_the_similar_reads}
Results of any $\oc$ operation is represented by a bit vector. 
Bit values of reads from the query set detected as similar to at least one read from the reference set are set to $1$ and the others are set to $0$.

\subsubsection{Compute logical operations on read subsets} 
\label{ssub:compute_logical_operations_on_read_subsets}
The bit vector representation is ideally suited to perform fundamental logical operations. 
\cp provides a module to perform the $AND$, $OR$ and $NOT$ operations between distinct subsets of a single initial set of reads. 

As presented in the simple case study (Section~\ref{ssec:simple_case_study}), these operations, although simple, are powerful while dealing with read subsets. 
They allow to combine comparison results and so to focus on read subsets intersections or exclusions. 

These logical operations perform very efficiently, both in terms of execution time and memory footprint. 
Moreover, it is worth to notice that they do not generate large result files, as results of these logical operations are also represented as bit vectors. 
This allows to intensively manipulate read subsets with no technical limitations.

\subsection{Dealing with more than two datasets}

We recall that the computation of the $A \oc B$ core operation involves indexation and search.
Once the $k$-mers of the reads from $B$ are indexed, then the $k$-mers of the reads from $A$ are sequentially search in the index. 
If more than a threshold number $t$ of such $k$-mers are find in the index, then the given read from $A$ is considered as similar to a read from $B$, which means that the associated value in the bit vector is set to $1$.

Consider $S=\{R_1, \dots, R_N \}$ a set of $N\geq2$ read sets. Applying \cp on the whole $S$ implies that $\forall(i,j) \in [1,N]^2, i<j$, three ordered operations are performed:
\begin{enumerate}
\item $R_i\oc R_j$
\item $R_j\ocf R_i = R_j\oc (R_i\oc R_j)$
\item $R_i\ocf R_j = R_i\oc (R_j\oc (R_i\oc R_j))$
\end{enumerate}
Note that for each couple $(i,j)$, the order $(i,j)$ or $(j,i)$ only slightly changes the overall results of the three operations.
To avoid redundancies, we limit these operations to $i<j$. 

\subsubsection{Factorizing the indexation} 

In practice, applying \cp on $S$ implies to perform the $R_i\oc R_j$ operations for all $i<j$. 
In particular, $R_1\oc R_N \dots R_{N-1}\oc R_N$ have to be computed. 
For these $N-1$ computations, the $k$-mer index of $R_N$ is the same. 
To avoid redundancies, the $R_N$ index is computed only once and the $N-1$ remaining sets are compared to $R_N$ using this single index. 
In general, while $R_{ref}$ ($ref\in [2,N]$) is indexed, the index is conserved in RAM memory during the computation of the $ref-1$ comparisons $R_{query}\oc R_{ref}$, with $query<ref$. 


\subsubsection{Results visualization} 
\label{sub:dendrograms_generation}
Comparisons of $N\geq2$ read sets $\{R_1, \dots , R_N \}$ provide useful metrics that give an overview of the genomic diversity of the studied samples. 
Those metrics are summarized in three matrices $M_1$, $M_2$, and $M_3$ with values calculated as follows:
\begin{itemize}
\item $M_1(i,j) = |R_i\ocf R_j|$ 
\item $M_2 (i,j) = 100 \times \frac{|R_i\ocf R_j|}{|R_i|}$
\item $M_3 (i,j) = 100 \times \frac{|R_i\ocf R_j|+|R_j\ocf R_i|}{|R_i|+|R_j|}$.
\end{itemize}

$M_1(i,j)$ with $(i,j) \in [1,N]^2$, is the raw number of reads from $R_i$ that are similar to at least one read from $R_j$.
As read sets may be of different sizes, $M_2(i,j)$ is the percentage of reads from $R_i$ similar to at least one read from $R_j$. 
Those two first matrices are asymmetrical.
$M_3$ is a symmetrical matrix. $M_3(i,j)$ is the percentage of similar reads between the two sets with respect to the total number of reads in $R_i$ and $R_j$.

For each matrix, a heatmap is generated .
Additionally, $M_3$ is used to construct a dendrogram representation by hierarchical clustering (see Fig~\ref{fig:heatmap_plain} for an example of a heatmap and a dendrogram generated by \cp). 

\subsection{The \cp modules}
\cp integrates four independent modules written in C++, all manipulating, as inputs and outputs, the bit vector representation of read subsets.
Additionally, \cp provides a python script (Commet.py) that takes $N\geq2$ read sets, filters them, compares them and generates explicit representations of comparative results, see Section~\ref{sub:automatically_filter_and_compare_ngeq2_read_sets}.

\subsubsection{Filtering and subsampling reads}
Thanks to the first module, $filter\_reads$, each read of each dataset (fasta or fastq format, gzipped or not) is filtered out according to user-defined criteria: minimal read length, number of undefined bases, and Shannon complexity~\cite{Shannon1948}, used to remove low complexity sequences. 
The result is a bit vector for each input read file. 
$Filter\_reads$ can also subsample each read set by limiting the number of selected reads to a user defined parameter $m$. 
The $m$ first reads that passed the filters are selected.

\subsubsection{Performing the $\oc$ core operation}
The second module, $index\_and\_search$, performs the $\oc$ core operation, representing results using the bit vector representation.
It inputs a set of read sets (the queries) to be searched in an indexed read set (the bank).
A read set may be composed of several read files. 
Each file could be associated to a bit vector.
In this latter case, $index\_and\_search$ only considers reads whose associated bit values are set to $1$.

%

\subsubsection{Manipulating read subsets}
The third module, $bvop$ (bit vector operations) inputs one or two bit vectors. 
In this second case, the two bit vectors should represent subsets of the same initial set. 
This module performs the $NOT$ operation on a single bit vector, and the $AND$, $OR$, and  $AND\ NOT$ operations on two bit vectors. 

\subsubsection{From bit vectors to read files}
Given an original read file and its bit vector, the last module, $extract\_reads$, generates an explicit representation of any read subset.

\subsection{Automatization for $N\geq2$ read sets}
\label{sub:automatically_filter_and_compare_ngeq2_read_sets}
\cp includes a python script (Commet.py) which inputs $N\geq2$ read sets. 
This pipeline i) filters reads, given user-defined parameters, ii) compares all-against-all read sets, and iii) outputs a user-friendly visualization of results. 
The outputs consist in the three matrices in $csv$ format, their heatmaps and a dendrogram as described in Section~\ref{sub:dendrograms_generation}. 
The dendrogram is realized using the \emph{hclust} \emph{R} function, computing a hierarchical complete clustering. 



\subsection{Combining read subsets use case} 
\label{ssec:simple_case_study}	
\begin{figure}[ht]
	\centering
		\includegraphics[width=.6\linewidth]{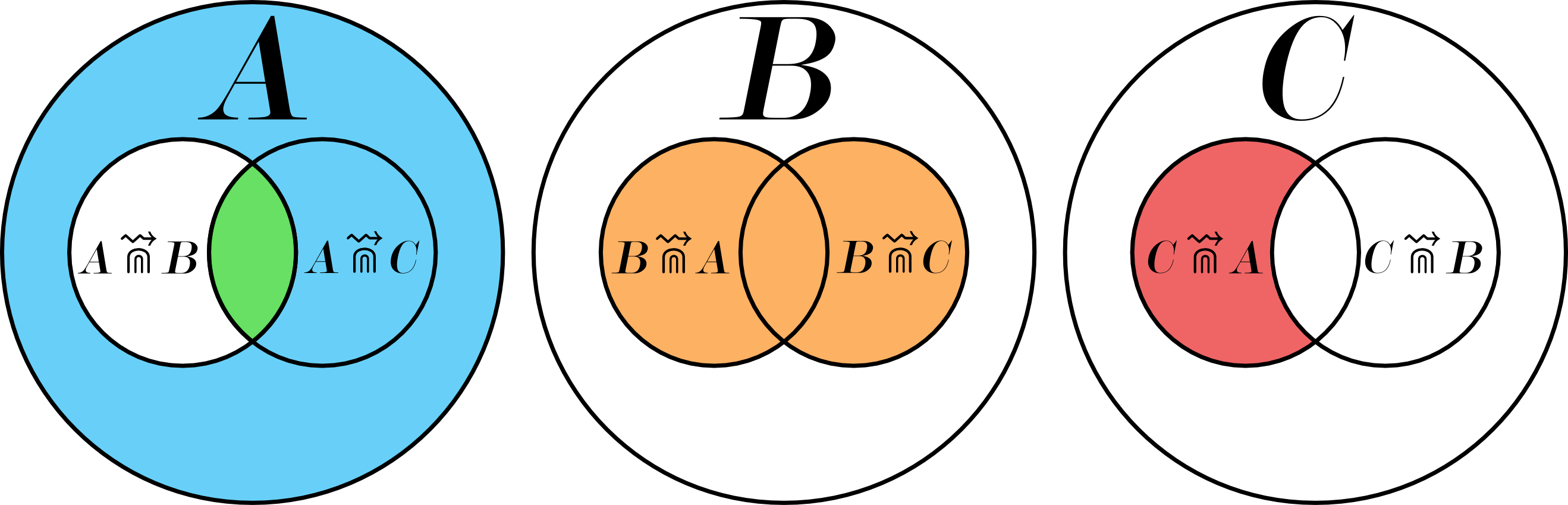}
		\caption{Logical operations on intersections between $A$, $B$ and $C$ extract read subsets of interest. The blue subset corresponds to the $A\ AND\ NOT(A\ocf B)$ operation. The green subset corresponds to the $(A\ocf B) \ AND\ (A\ocf C)$ operation. The orange subset corresponds to the $(B\ocf A)\ OR\ (B\ocf C)$ operation. The red subset corresponds to the $(C\ocf A) \ AND\ NOT(C\ocf B)$ operation.}
	\label{fig:subsets}
\end{figure}
%
%

By using the $bvop$ module, logical operations can be performed between inputs/outputs of the \cp pipeline output. 
For instance, reads from $A$ not similar to any read from set $B$ (blue subset of Fig~\ref{fig:subsets}) are obtained by first applying $NOT(A\ocf B)$ operation.
Reads from $A$ similar to at least one read from $B$ and one read from $C$ (green subset of Fig~\ref{fig:subsets}), are identified by computing the $AND$ operation: $(A\ocf B) \ AND\ (A\ocf C)$. 
In the same spirit, reads from $B$ similar to at least one read from $A$ or one read from $C$ (orange subset of Fig~\ref{fig:subsets}), are found by computing the $OR$ operation: $(B\ocf A)\ OR\ (B\ocf C)$. 
Operations may be combined to obtain more complex results as, for instance, the red subset of Fig~\ref{fig:subsets}, representing reads from $C$ similar to at least one read from $A$, but not similar to any read from $B$. 
This would be done by applying the $(C\ocf A) \ AND\ NOT(C\ocf B)$ operation.

\section{Results} 
\label{sec:results}

\subsection{\cp efficiently compares multiple metagenomes} 
\label{sub:gut_big}

\begin{figure*}[t]
	\centering
	\begin{tabular}{cc}
		\includegraphics[width=0.30\linewidth]{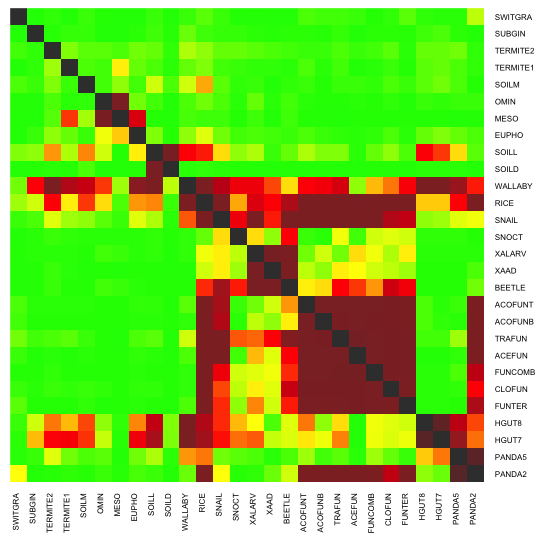} & \includegraphics[width=0.35\linewidth]{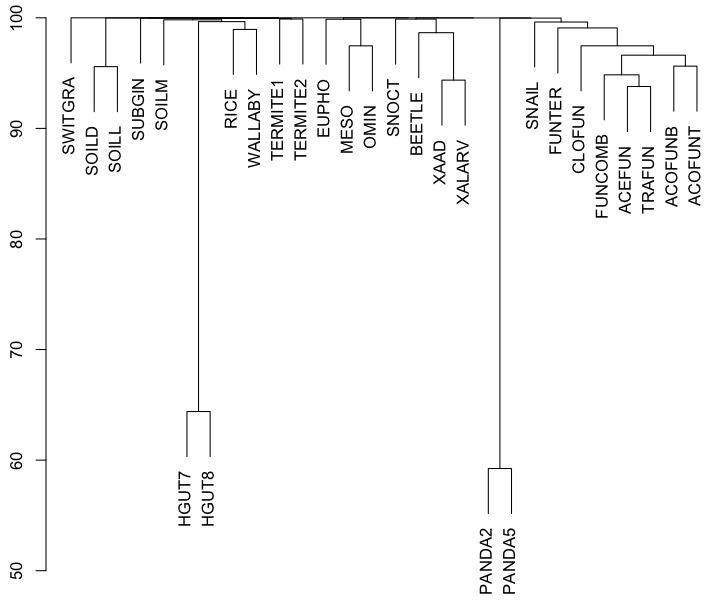}
	\end{tabular}
	\caption{Heatmap (left) and dendrogram (right) representation of the results of the comparison of 28 datasets from the IMG/M database. Results are given with $t=2$, $m=10000$ and $k=33$. The heatmap is constructed from the matrix $M_2$ and is thus asymmetrical. The dendrogram is constructed from the matrix $M_3$ by the hierarchical clustering procedure available in \emph{R} (method ``complete'').}
	\label{fig:heatmap_plain}
\end{figure*}

%

\begin{table}[t]
\caption{28 metagenomes from the IMG/M database}
\begin{center}
	\tiny
\begin{tabular}{ll}
Identifiers & Description \\
\hline
SWITGRA & Rhizosphere soil from \textit{Panicum virgatum} \\
SUBGIN & Oral TM7 microbial community of Human \\
TERMITE2, TERMITE1 & Gut microbiome of divers termites \\
SOILM, SOILD, SOILL & Soil microbiome from divers locations\\
OMIN, MESO, EUPHO & Divers marine planktonic communities\\
BEETLE & \textit{Dendroctonus ponderosae} \\
ACOFUNT, ACOFUNB, CLOFUN, & Fungus garden of divers ants \\
ACEFUN, TRAFUN, FUNCOMB & \\
FUNTER & Fungus-growing termite worker\\ 
WALLABY & Forestomach microbiome of tammar wallaby\\
RICE & Endophytic microbiome from rice\\
SNAIL & \textit{Achatina fulica} \\
SNOCT &  \textit{Sirex noctilio} microbiome\\
XALARV, XAAD & \textit{Xyleborus affinis} microbiome (larvae, adult)\\
HGUT7, HGUT8 & Human gut community\\
PANDA2, PANDA5 & Wild panda gut microbiome\\
\end{tabular}
\end{center}
\label{tab:microbiom}
\end{table}%

We tested \cp on a set of 28 metagenomes from the IMG/M database~\cite{Markowitz2012} (see Table \ref{tab:microbiom}).
These 28 metagenomes were compared with options $k=33$, $t=2$ and $m=10000$.
Computations were done using \cp (Commet.py) and Compareads (v1.3.1) on a 2.9 GHz Intel Core i7 processor with 8GB of RAM and a Solid-State Drive.
\cp calculated the 756 intersections in 35 minutes while Compareads took 81 minutes.
In this experiment, \cp is 2x faster than Compareads thanks to its indexing strategy (each file is fully indexed only once).
The obtained dendrograms, shown in Figure~\ref{fig:heatmap_plain}, are biologically coherent.
The different fungus samples are grouped together as well as soil samples, marine planktonic communities and insects.
The two human gut microbiome samples are far from other species, as well as the two panda gut microbiome samples.

\subsection{Metasoil study} 
\label{sub:metasoil}
\begin{figure}[ht]
	\centering
	\begin{tabular}{c}
		\includegraphics[width=.77\linewidth]{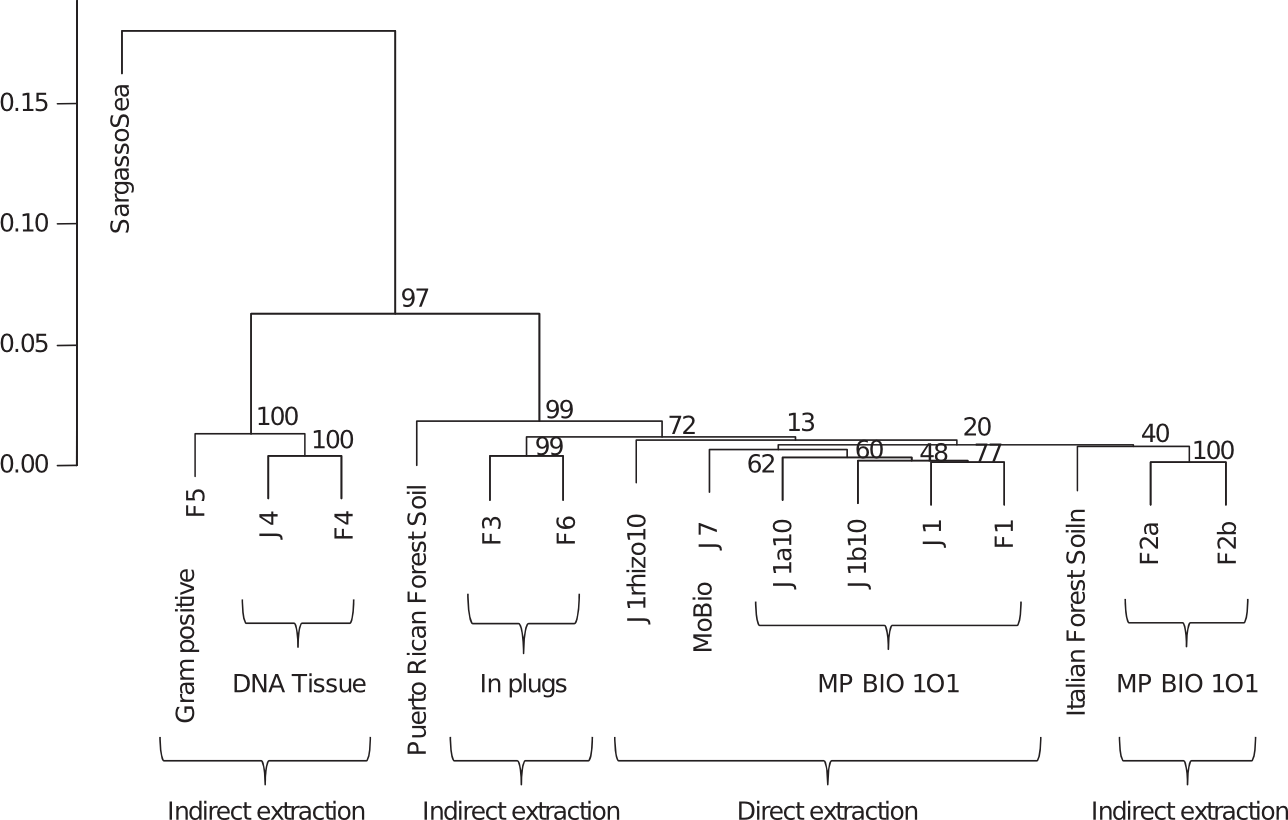}
		\\
		\includegraphics[width=.77\linewidth]{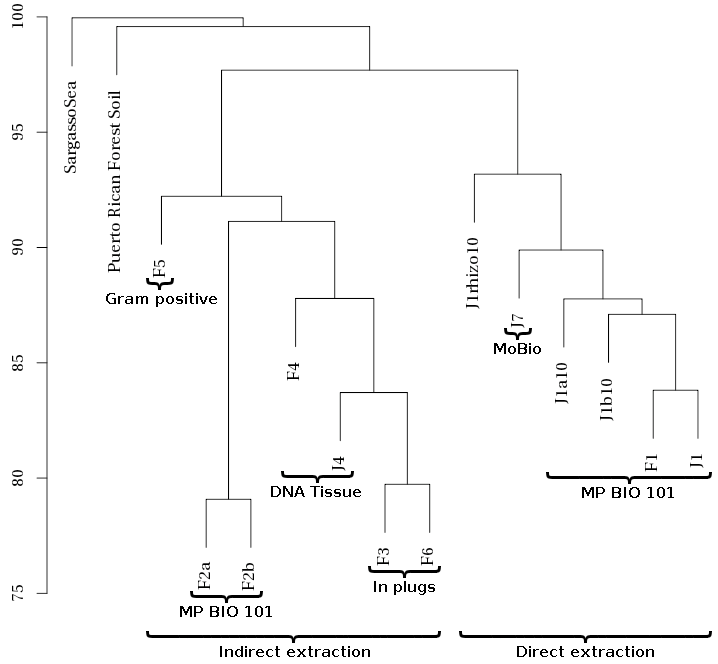}
	\end{tabular}
	\caption{Dendrograms from MetaSoil study (top, figure from~\cite{Delmont2012}) and \cp analysis (bottom), comparing the 13 MetaSoil samples, an other soil metagenome and a seawater metagenome (Sargasso Sea).}
	\label{fig:metasoil}
\end{figure}

The MetaSoil study focuses on untreated soils of Park Grass Experiment, Rothamsted Research, Hertfordshire, UK. 
One of the goals of this study is to assess the influence of depth, seasons and extraction procedure on the sequencing~\cite{metasoil1}. 
To achieve this, the 13 metagenomes from MetaSoil, two other soil metagenomes and a sea water metagenome, were compared at the functional level using MG-RAST~\cite{mgrast}. 
This approach identified 835 functional subsystems present in at least one of those metagenomes. 
On Figure~\ref{fig:metasoil}.a, samples were clustered using the relative number of reads associated with the 835 functions. 
This figure shows that the extraction procedure correlates with sample clusters: two metagenomic samples processed with the same extraction procedure share more similarities at the functional level than two samples processed with different extraction procedures~\cite{Delmont2012}.

This study was reproduced with \cp on all available metagenomes.
The generated bit vectors weigh 68MB while the explicit representation of the fasta results requires 6.4GB. The storage footprint is thus divided by a factor 100. This ratio is even higher if using fastq format or if dealing with larger read files. The \cp computation time was 828 minutes (the same set treated by Compareads took 2981 minutes).

Although \cp uses another metric, the produced dendrogram is highly similar to the MetaSoil one (see Fig~\ref{fig:metasoil}).
On both dendrograms, samples coming from direct extraction are clustered together and external metagenomes are far from the MetaSoil's.
Moreover, on the \cp dendrogram, all samples coming from indirect extraction are clustered together, which is not the case in the MetaSoil study.
Even if the two comparing methods are different, they lead to the same conclusion: extraction procedures have a critical impact on sequencing. 

\section{Conclusion}
\cp gives a global similarity overview of all datasets of a large metagenomic project.
It performs all-against-all comparisons of $N$ datasets by factorizing indexation phases. 
Disk I/Os and storage footprint are highly limited thanks to a new read subset representation which reduces the storage space by at least two orders of magnitude compare to explicit fasta or fastq format. 
Interestingly, this read subset representation is a powerful way to compute extremely fast boolean operations between read subsets without copying large read files. 
This enables to focus on reads that fulfill several distinct constraints of interest. 
The advantages of this representation and of the boolean manipulation are not limited to the \cp framework. 
Thus, \cp includes a C++ library of reusable components to manipulate read subsets. 

\cp produces graphical outputs that sum up all-against-all comparisons results and open the way for further statistical analysis, thanks to the provided similarity matrices.


A future work will consists in quickly identify significant clusters of read sets by applying rougher comparative metrics (such as the GC content) or a statistical framework based on Principal Component Analysis (PCA). Then, \cp should be used to go further by precisely compute the shared reads between read sets inside clusters, or between clusters.

\cp is available under the A-GPL license: http://github.com/pierrepeterlongo/commet.

\section*{Acknowledgment}
Authors warmly thank Claire Lemaitre for her precious advices and her help designing the \emph{R} functions.
This work was supported by the french ANR-2010-COSI-004 \emph{MAPPI} and by the ANR-12-BS02-0008 \emph{Colib’read} projects.
Guillaume Collet's work is funded by the investment expenditure program IDEALG 1192 ANR-10-BTBR-02-04-11.



\bibliographystyle{IEEEtran}
\bibliography{IEEEabrv,commet}
%
%
%

\end{document}